\documentclass[reprint, superscriptaddress, amsmath, amssymb, aps, pra]{revtex4-2}

\usepackage{graphicx}
\usepackage{dcolumn}
\usepackage{bm}
\usepackage{xcolor}
\usepackage{soul}
\usepackage{pdfcomment}
\usepackage[normalem]{ulem}

\begin{document}

\preprint{APS/123-QED}

\title{\textbf{Gate Dispersive Charge Detection in Planar Ge/SiGe Quantum Dots} 
}%

\author{M.~R.~Tanvir}
\affiliation{Department of Electrical and Computer Engineering,
The University of British Columbia,
2332 Main Mall, Vancouver, British Columbia V6T~1Z4, Canada}
\affiliation{Stewart Blusson Quantum Matter Institute, University of British Columbia, Vancouver, BC, Canada}

\author{E.~Sajadi}
\affiliation{Department of Electrical and Computer Engineering,
The University of British Columbia,
2332 Main Mall, Vancouver, British Columbia V6T~1Z4, Canada}
\affiliation{Stewart Blusson Quantum Matter Institute, University of British Columbia, Vancouver, BC, Canada}
\affiliation{Department of Physics and Astronomy, University of British Columbia, Vancouver, BC, Canada}

\author{M.~Veldhorst}
\affiliation{QuTech and Kavli Institute of Nanoscience, Delft University of Technology, P.O. Box 5046, 2600 GA Delft, The Netherlands}

\author{G.~Scappucci}
\affiliation{QuTech and Kavli Institute of Nanoscience, Delft University of Technology, P.O. Box 5046, 2600 GA Delft, The Netherlands}

\author{J.~Salfi}
\email[Corresponding author: ]{jsalfi@ece.ubc.ca}
\affiliation{Department of Electrical and Computer Engineering,
The University of British Columbia,
2332 Main Mall, Vancouver, British Columbia V6T~1Z4, Canada}
\affiliation{Stewart Blusson Quantum Matter Institute, University of British Columbia, Vancouver, BC, Canada}
\affiliation{Department of Physics and Astronomy, University of British Columbia, Vancouver, BC, Canada}

\date{\today}

\begin{abstract}
Spin qubits are an attractive platform for scalable quantum computation, and recently, laterally gated quantum dots (QDs) in Ge heterostructures have emerged as a leading implementation. Radio-frequency (RF) single-hole transistor (SHT) spin readout and integrated superconducting resonator charge readout have demonstrated in Ge QDs, but their wiring overhead and on-chip size can pose challenges, while gate dispersive (GD) readout, which requires less on-chip overhead, has yet to be demonstrated for Ge QDs. Here we demonstrate compact, low-overhead RF-GD charge readout of a Ge double QD (DQD), both directly, and assisted by a proximal single-hole box (SHB). We resolve QD loading and interdot transitions with both techniques. We obtain unity signal-to-noise ratio (SNR) in 2.7 millisecond and 135 microsecond integration time, for gate-dispersive detection of the DQD and SHB-DQD systems, respectively. We find that the SHB enhances the interdot signal-to-noise ratio (SNR) of our 540 MHz circuit by a factor of 20 for a DQD coupling $t_c \approx 5$ GHz. Power and temperature dependent measurements verify that SNR is limited by dielectric loss. We argue that with achievable improvements in dielectric loss and amplification, high-fidelity (99\%) readout in less than 1 microsecond should be possible, assuming ideal spin-to-charge conversion, which has seen rapid recent progress. These results establish a simple and compact gate-based readout approach for direct and SHB-enhanced readout of Ge heterostructure QDs.

\end{abstract}

\maketitle

Spin qubits are a promising platform for scalable quantum computation because they offer long coherence times and high-temperature operation with small footprint and compatibility with advanced semiconductor manufacturing, which could enable high-density integration \cite{lossQuantumComputationQuantum1998, vandersypen2017interfacing, camenzind2022hole, burkardSemiconductorSpinQubits2023, huang2024high}. Group-IV laterally-gated quantum dots (QDs)\cite{lossQuantumComputationQuantum1998} are particularly attractive owing to their low natural abundance of spin-bearing nuclei and potential for isotopic enrichment, where the latter virtually eliminates nuclear decoherence \cite{tyryshkin2012electron, muhonen2014storing, veldhorst2014addressable, kobayashiEngineeringLongSpin2021}. Hole spins in planar silicon MOS, planar Ge heterostructures, and silicon nanowire QDs benefit from intrinsic spin-orbit interactions, which enable all-electric spin manipulation \cite{maurand2016cmos,scappucciGermaniumQuantumInformation2020, hendrickxFourqubitGermaniumQuantum2021, jirovecSinglettripletHoleSpin2021, vorreiter2026precision}, and lack orbital/valley degeneracy, which complicates spin-based devices \cite{goswami2007controllable, yang2013spin, van2014probing, van2018readout, voisin2020valley}. Planar Ge heterostructure QDs also benefit from low disorder and light effective mass\cite{lodariLightEffectiveHole2019}, simplifying device fabrication and enabling rapid progress on coherent control of individual holes \cite{hendrickxSingleholeSpinQubit2020}, high-fidelity two-qubit operations \cite{hendrickxFastTwoqubitLogic2020}, connected arrays of QDs \cite{borsoiSharedControl162024, wangOperatingSemiconductorQuantum2024, kelly2026identification,john2025robust, seidlerSpatialUniformityGtensor2025}. Recently, isotope purification has been reported in Ge heterostructures and QDs\cite{moutanabbir2024nuclear,daoust2026nuclear,zeng2026high}. As Ge QDs scale in complexity, compact circuits for fast, high-fidelity readout become increasingly important \cite{vandersypenInterfacingSpinQubits2017,zhengRapidGatebasedSpin2019, borjansSpinDigitizerHighfidelity2021}.

Spin readout in semiconductor QDs is performed in two steps, spin-to-charge conversion via e.g. spin-dependent tunneling (i.e., Pauli spin blockade (PSB) \cite{ono2002current}) or energy-selective tunneling, and detection of the resulting charge state\cite{elzermanSingleshotReadoutIndividual2004, pettaCoherentManipulationCoupled2005, hansonSpinsFewelectronQuantum2007, morelloSingleshotReadoutElectron2010}. Radio-frequency (RF) reflectometry \cite{schoelkopfRadioFrequencySingleElectronTransistor1998} has enabled high-bandwidth charge detection with, e.g., single-electron transistors (SETs), and has been used for single-shot spin readout by PSB. In laterally gated Si/SiGe devices, a readout fidelity $F\approx99.9$\% has been reported for a measurement time $t_m=2$~$\mu$s \cite{takedaRapidSingleshotParity2024}, and in Si-MOS, $F\approx 99.34$\% and $t_m=50~\mu$s\cite{huang2024high}. With RF-SHTs in planar Ge heterostructures, single-shot hole-spin readout with $t_m\approx 5$~$\mu$s was performed, and near-ideal spin-to-charge conversion ($F\approx 97$~\%) has recently been achieved in $t_m=52$~$\mu s$\cite{hendrickxSingleholeSpinQubit2020, vukusicSingleShotReadoutHole2018, wangOperatingSemiconductorQuantum2024, kelly2026identification,john2025robust}. A disadvantage of RF-SET/SHT readout is the additional wiring and device overhead: the sensor requires its own QD, two reservoirs, and electrical connections \cite{vandersypenInterfacingSpinQubits2017, kunneSpinBusArchitectureScaling2024}. 

RF gate dispersive (GD) detection, where the QD gate is directly connected to the RF impedance-matching circuit\cite{collessDispersiveReadoutFewElectron2013, gonzalez-zalbaProbingLimitsGatebased2015, rossiDispersiveReadoutSilicon2017, westGatebasedSingleshotReadout2019, ahmedRadioFrequencyCapacitiveGateBased2018, scarlinoAllMicrowaveControlDispersive2019}, offers low-overhead (compact) readout. Changes in QD admittance modify the amplitude and phase of a reflected RF signal, allowing charge transitions to be detected without a separate sensor \cite{ansaloniGateReflectometryDense2023, mizutaQuantumTunnelingCapacitance2017, svabDirectDispersiveSignature2026}. The technique was extensively applied to QDs in GaAs, silicon, Si/SiGe heterostructures, and silicon nanowires \cite{collessDispersiveReadoutFewElectron2013,betzDispersivelyDetectedPauli2015, van2018readout, crippaGatereflectometryDispersiveReadout2019, ciriano-tejelSpinReadoutCMOS2021, demichielisSiliconSpinQubits2023, denisovDispersiveReadoutSilicon2023}. RF-GD readout can be slower than RF-SET readout since the signal is proportional to the lever-arm difference between the QDs, c.f. the Coulomb repulsion difference between the QDs and sensor, for RF-SET. In planar devices, especially buried quantum wells, the inter-dot lever-arm differences can limit readout speed. Previous work demonstrated $F\approx 74.5$\% with $t_m=2.6~\textrm{ms}$ in a planar Si-MOS device \cite{westGatebasedSingleshotReadout2019}, and $F\approx90$\% with $t_m=100~\mu\textrm{s}$ in a Si nanowire \cite{hamonicFoundryfabricatedSpinQubit2025}.  While GD readout has been demonstrated in Si/SiGe, Si/SiO2 and GaAs QDs, Ge materials have a more complex buffer layer structure\cite{lodariLightEffectiveHole2019} which is often has more RF loss in comparison. Recent studies have established integration of high Q superconducting resonators with Ge substrates and QDs\cite{kangCouplingHoleDouble2024, depalmaStrongHolephotonCoupling2024,ruggiero2026high}. Despite the advantages of \textit{in-situ} RF-GD readout, it has yet to be reported for planar Ge hole QDs.

The performance of RF-GD spin readout has recently been improved using auxiliary QDs (a single electron box (SEB)) with a single lead (instead of two leads for an SET)\cite{houseHighSensitivityChargeDetection2016, borjansSpinDigitizerHighfidelity2021,niegemannParitySingletTripletHighFidelity2022, oakesFastHighFidelitySingleShot2023, cochraneIntrinsicNoiseSingleelectron2024}. The mutual Coulomb repulsion between the SEB and the QDs converts a change an interdot DQD charge into a change in SEB charge whose signal is proportional to the difference in the mutual Coulomb repulsion of each QD to the SEB, c.f. the difference in lever arm of each QD to the gate, for pure GD readout \cite{gonzalez-zalbaProbingLimitsGatebased2015}. The performance of RF-SEBs, comparable to RF-SETs, has been reported: $F\approx99$\% with $t_m=20~\mu\textrm{s}$ in Si-MOS \cite{laineHighfidelityDispersiveSpin2026}, and $F\approx 99.2$\% with $t_m=100~\mu\textrm{s}$ in Si/SiGe \cite{borjansSpinDigitizerHighfidelity2021}. When the mutual Coulomb repulsion of the QDs to the SEB is strong enough, a cascade readout \cite{vandiepenElectronCascadeDistant2021} can enhance the RF-gate dispersive (GD) signal\cite{chittock-woodRadiofrequencyCascadeReadout2026}. The performance of gate-based readout has been demonstrated in Si/SiGe using higher frequency on-chip superconducting resonators, with higher Q achieved via weak capacitive coupling providing $F\approx 98\%$ with $t_m=6~\mu\textrm{s}$ \cite{zhengRapidGatebasedSpin2019}.

Here, we demonstrate RF-GD charge readout of a few-hole DQD operating at $540$~MHz in planar Ge/SiGe QD system. We first resolve the \textit{in-situ} GD response and then identify a coupled auxiliary QD that can be tuned for GD SHB charge readout. In our device, at the $(1,1)$--$(2,0)$ interdot transition, we achieve a power-SNR of $18.5\pm3$ with $t_m=50~\textrm{ms}$ for the \textit{in-situ} GD signal, while the SHB configuration produces an approximately twenty-fold enhancement in power-SNR, $372\pm12$ for the same $t_m=50~\textrm{ms}$. This enhancement arises because the interdot charge transition produces a larger differential change in the Coulomb interaction with the SHB than the corresponding change in quantum capacitance for the \textit{in-situ} GD signal. Using our measurements, we identify a set of achievable conditions for the readout circuit to achieve high-fidelity spin readout in less than 1 microsecond. 

\begin{figure*}
    \centering
    \includegraphics{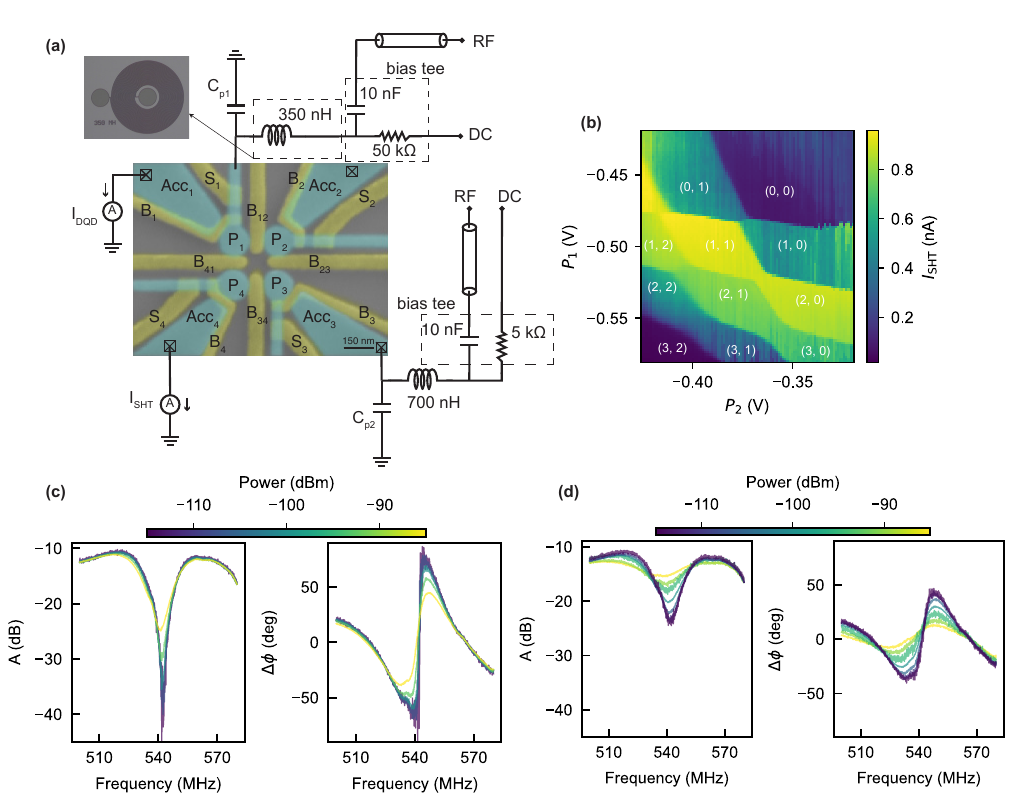}
    \caption{Measurement circuit and characterization of the RF matching circuits. (a) Schematic of the measurement setup together with a false-colored scanning electron micrograph (SEM) image of the Ge/SiGe QD device with lateral gates in a lower layer (yellow) and upper layer (teal) separated by gate dielectric. Superconducting NbTiN spiral inductors located on separate Si chips are connected via wirebonds to the QD chip: a 350 nH inductor is connected to plunger gate $P_1$ for GD readout (optical micrograph shown in the inset) and a 700 nH inductor is connected to an ohmic contact to implement a conventional RF single-hole transistor (RF-SHT) not characterized here. (b) SHT current as a function of plungers $P_1$ and $P_2$ with gate voltage $P_3$ compensated for SHT coupling to $P_1$ and $P_2$. (c) Amplitude and phase response of the RF-GD matching circuit, exhibiting a 540 MHz resonance. (d) Same as c, but with a 0.5 $\mu$S conductance channel beneath gates $\textrm{Acc}_1$, $P_1$, $P_2$, and $\textrm{Acc}_2$. For both c and d, increasing the applied RF power reduces the resonance depth and phase response, indicating increased dissipation causing further departure from the impedance-matching condition.}
    \label{fig:setup_and_tank_response}
\end{figure*}

Fig.~\ref{fig:setup_and_tank_response}(a) is a schematic of the QD device and measurement circuit. The device contains two layers of gates: inter-QD barriers ($B_{ij}$), reservoir barriers ($B_i$), screening gates ($S_i$), QD plungers ($P_i$), and reservoir accumulation gates (${\rm Acc}_i$). The LC matching circuit for RF-GD reflectometry employs a 350 nH off-chip superconducting (NbTiN) spiral inductor, wirebonded to the bond-pad for gate $P_1$, with no intentional capacitor. The LC matching circuit for RF-SHT reflectometry employs an off-chip 700 nH NbTiN sprial inductor wirebonded to the ohmic under the accumulation gate $\rm{Acc}_3$, with a surface-mount capacitor. Reflection measurements are performed by high-electron-mobility transistor (HEMT)-based low-noise amplifiers at 4~K connected to the spiral inductors via directional couplers. Current pre-amplifiers are connected to the ohmics under reservoir gates $\rm{Acc}_1$ and $\rm{Acc}_4$. Further details of the device are supplied in Appendix \ref{appendix: Heterostructure and device fabrication}.

The device is tuned to a regime with a DQD under plungers $P_1$ and $P_2$ and an SHT under plungers $P_3$ and $P_4$ [Fig.~\ref{fig:setup_and_tank_response}(a)], at the base temperature of a dilution refrigerator. To start, two isolated channels with conductance around $0.5$~$\mu$S are formed between $\textrm{Acc}_1$ and $\textrm{Acc}_2$ and between $\textrm{Acc}_3$ and $\textrm{Acc}_4$ using gate voltages, with $B_{23}$ and $B_{41}$ voltages held to values that electrically isolate the channels. Refining the gate voltages yields an SHT on the bottom channel and a few-hole DQD on the top channel. Screening gates ({$S_i$}) are held at 0~V during this initial tuning procedure. Once the SHT and DQD are independently set up, the DQD ohmic voltages are turned off, the SHT is tuned to the shoulder of a Coulomb oscillation so that it acts as a charge sensor, and the DQD charge occupation is monitored by the SHT current as a function of voltages $P_1$ and $P_2$. A stability diagram of the few-hole DQD measured by the SHT is presented in Fig.~\ref{fig:setup_and_tank_response}(b). 

The RF-GD sensor was characterized during the tuning procedure: the resonance with all gates at 0~V [Fig.~\ref{fig:setup_and_tank_response}(c)] occurs at $f_0\approx 540$~MHz. The dip in reflection of approximately 30 dB reveals the circuit is closely impedance matched. Using $L=350$~nH we deduce a circuit capacitance of $C\approx 0.25$~pF and optimal matching condition $R=L/CZ_0\approx 28$~k$\Omega$, and a loss of $R_\varepsilon\approx 25k$~k$\Omega$ close to the matching condition (see Fig.~\ref{fig:CircuitModel_SNR}(b) in Appendix~\ref{appendix: comments on high-fidelity single-shot readout}). When the channel with $P_1$ and $P_2$ is tuned to a conductance 0.5~$\mu$S, the  dip reduces to $\approx 12$~dB [Fig.~\ref{fig:setup_and_tank_response}(d)], and we estimate $R_\varepsilon\approx 18$~k$\Omega$. As the applied RF power on the RF-GD sensor increases, as presented in Figs.~\ref{fig:setup_and_tank_response}(c,d), the resonance dip becomes somewhat shallower, i.e., matching is worsened. With holes accumulated under SHT, the dip reduces to $\approx 5$~dB (Appendix~\ref{appendix: Additional data}), with $R_\varepsilon\approx 8$~k$\Omega$. The RF-GD circuit, now set up, is used to characterize the device together with current measurements on the SHT. 

\begin{figure*}
    \centering
    \includegraphics{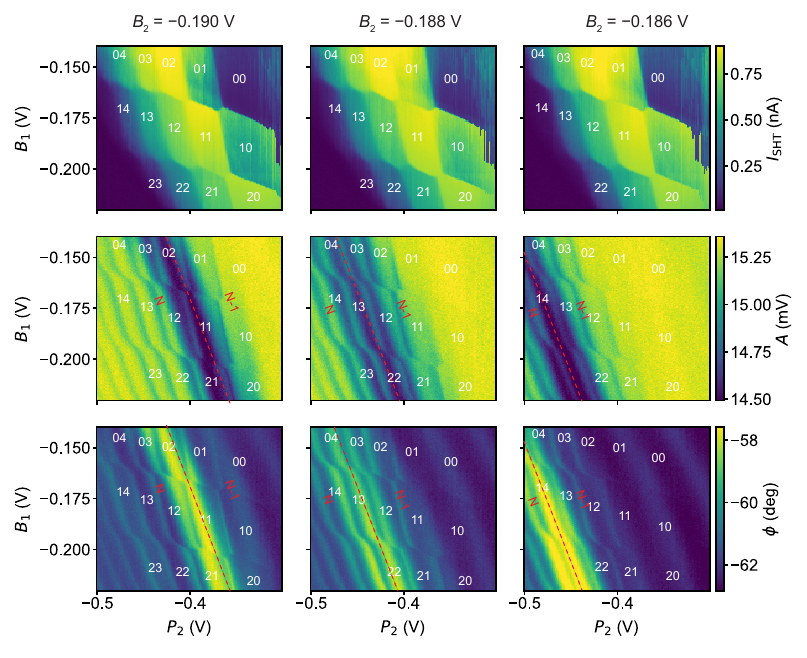}
    \caption{Evolution of the DQD stability diagram as the SHB is tuned by gate $B_2$. Each column corresponds to a different value of $B_2$, while the rows show the SHT current (top), the amplitude (middle), and phase (bottom) of the gate-attached resonator. The SHT plunger gate is compensated to stay tuned onto the shoulder of a Coulomb oscillation. Charge states are labelled with the notation $nm$, $n$ holes in ${\rm QD}_1$ and $m$ holes in ${\rm QD}_2$, using the SHT stability diagram. The SHB charge transition (diagonal red line) is shifted across the stability diagram by varying $B_2$ and allows detection of the DQD charge. QD$_2$ loading peaks and some interdot peaks are resolved in the RF-GD response, and some interdot and loading transitions are captured by the RF-GD-SHB response.}
    \label{fig:moving_stripes}
\end{figure*}

We now investigate the RF-GD response~\cite{collessDispersiveReadoutFewElectron2013} when the device is tuned to operate as an SHT coupled to a few-hole DQD. The response to $P_1$ and $P_2$ gates (Appendix \ref{appendix: Additional data}) reveals the honeycomb DQD pattern matching the charge sensor response, but with additional localized (QD) states strongly coupled to $P_1$. The additional localized (QD-like) state does not appear in measured currents, has negligible coupling to other gates, and does not interact with (shift the load-lines of) the DQD. A second (auxiliary) localized (QD) state, which does not appear in measured currents, is also detected by the RF-GD circuit. Our detailed characterization shows that the second auxiliary QD couples to $B_{12}$, $B_2$, $S_1$, $P_1$, and $P_2$ with relative lever arms of $1.0,\ 0.54,\ 0.11,\ 0.09,\ \textrm{and}\ 0.04$, respectively. We refer to the second localized QD state as an SHB, because it only couples to one lead, i.e., it is not detected in current measurements.

We now show the effectiveness of the RF-GD circuit in detecting directly, and through the SHB, the charge state of the DQD. We accomplish this by measuring the RF-GD reflection dependence on gates $B_1$ and $P_2$ with the RF excitation still on $P_1$. Figure~\ref{fig:moving_stripes} presents SHT-measured and RF-GD-measured response for three different $B_2$ voltages, increasing from left to right in $2$~mV increments. The SHT resolves the DQD. Peaked resonant features are observed in the RF-GD response that coincide with $\textrm{QD}_2$ load lines for the SHT. These QD$_2$ load-lines sharpen and eventually disappear as $P_2$ increases (becomes less negative) in Fig.~\ref{fig:moving_stripes}, as expected, as reservoir-$\textrm{QD}_2$ coupling reduces and falls below the GD resonant frequency~\cite{houseRadioFrequencyMeasurements2015}. QD$_1$ load lines do not appear. Indeed, QD$_1$ has a weaker reservoir coupling than QD$_2$, as evidenced by the sharper QD$_1$ load lines measured by the SHT. We additionally observe the SHB resonance, which shifts to lower $P_2$ values as $B_2$ increases, and shifts appreciably when it intersects the DQD loading and interdot transitions, similar to how our SHT measures charge. Charge numbers are indicated on the stability diagram. As discussed in the appendix \ref{appendix: Constant-interaction model}, the chemical potential shift of the SHB is too low for cascade readout.

\begin{figure}
    \centering
    \includegraphics{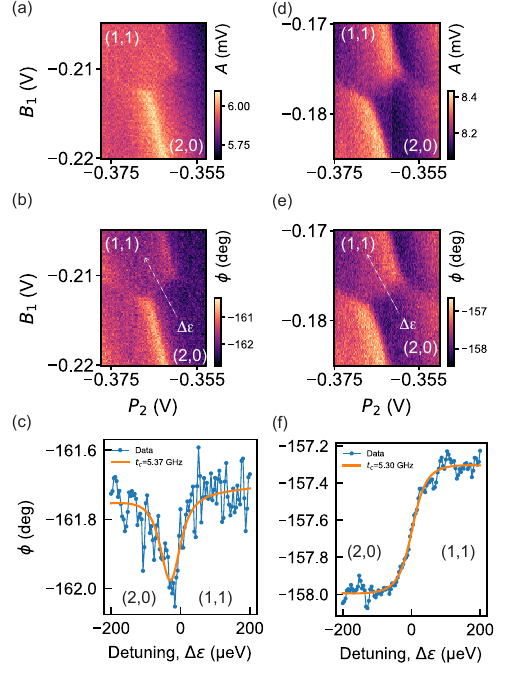}
    \caption{Comparison of the RF-GD and RF-GD-SHB response for -95~dBm at the device. (a),(b) RF-GD amplitude and phase response near the $(1,1)$--$(2,0)$ interdot charge transition with SHB off-resonance. (c) Phase response (average of 3 traces) along the detuning axis in (b), with the characteristic quantum-capacitance dip used to extract the interdot tunnel coupling. (d),(e) Same as (a),(b) but with the RF-GD-SHB on-resonance, and operating as a charge sensor (f) Phase response (average of 3 traces) along the detuning axis in (e), with the broadened step response for the SHB. The SHB improves the SNR by a factor of approximately 20, from 18.5 to 372. We verified the noise in panels (c) and (f) to be the amplifier white noise.}
    \label{fig:two_modes}
\end{figure}

Next, we extract the SNR for the RF-GD and RF-GD-SHB charge response at the $(1,1)$--$(2,0)$ interdot transition relevant to spin readout by PSB, and establish a 20x improvement for the latter. Figure~\ref{fig:two_modes} compares the RF-GD response obtained when the auxiliary QD is tuned away from the DQD interdot transition [Figs.~\ref{fig:two_modes}(a,b)] to the response when the SHB resonance overlaps the interdot transition [Figs.~\ref{fig:two_modes}(c,d)]. First, with the auxiliary QD tuned away from the interdot transition, a narrow dip is observed at zero detuning [Fig.~\ref{fig:two_modes}(c)]. The line cut is obtained by averaging three traces, corresponding to an effective integration time of $50~\textrm{ms}$. Fitting the phase to a model that takes into account interdot tunneling and temperature (Appendix \ref{appendix: tunnel coupling in direct regime}), we find a tunnel coupling $t_c\sim5.37\pm0.82$ GHz and power-SNR \cite{zhengRapidGatebasedSpin2019} of $18.5\pm3$. An integration time of about $2.7~\textrm{ms}$ would yield a unity-SNR. Figures~3(d,e) show the phase response when the SHB resonance overlaps the interdot transition. The line cut of phase along the detuning axis [Fig. 3(f)]---which is also obtained by averaging three traces, corresponding to an effective $t_m = 50~\textrm{ms}$---contains a broadened step response, as expected from an electrochemical shift, induced by the inter-QD charge transition, that is small compared to the broadening of the SHB resonant peak. Fitting the phase to a model taking into account tunnel coupling and temperature (Appendix \ref{appendix: tunnel coupling in SHB regime}), we obtain $t_c\sim 5.23\pm0.70$ GHz and a power-SNR of $372\pm12$ for $t_m = 50~\textrm{ms}$. This yields a minimum integration time of $135~\mu\textrm{s}$ for an SNR of 1, 20x faster than the in-situ RF-GD. 

\begin{figure*}
    \centering
    \includegraphics[width=\linewidth]{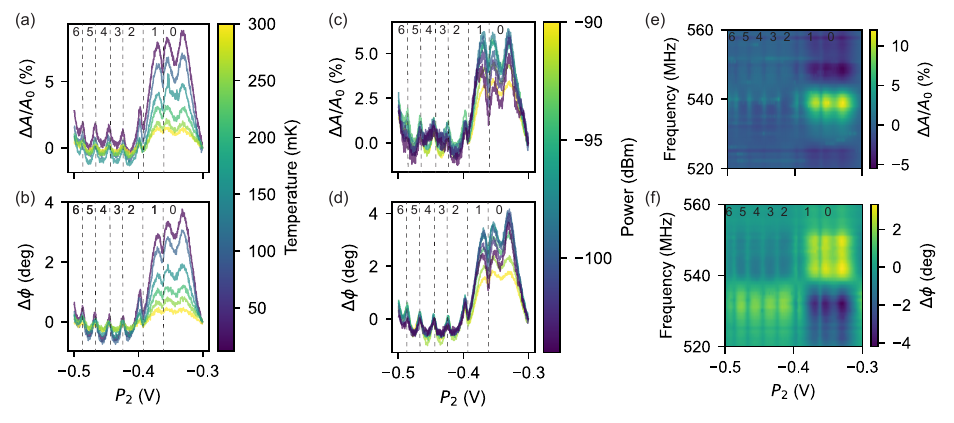}
    \caption{Dependence of the GD readout on temperature, RF excitation power, and probe frequency. All measurements are performed along a line cut through the charge stability diagram at $B_1=-0.18~\mathrm{V}$ (shown in Appendix Fig.~\ref{fig:stability_diagram_for_line_cuts}), containing signatures of both the in-situ GD readout of the reservoir-$\textrm{QD}_2$ charge transitions and the auxiliary-dot-assisted readout. The ${\rm QD}_2$ occupation number $n_2$ is labeled. (a,b) Reflected amplitude and phase as a function of temperature at -95~dBm power. (c,d) Dependence of the reflected amplitude and phase on the applied RF excitation power. (e,f) Dependence of the reflected amplitude and phase on probe frequency.}
    \label{fig:dependence}
\end{figure*}

We now investigate in Fig.~\ref{fig:dependence} how temperature, RF power, and probe frequency impact the RF-GD and RF-GD-SHB readout, along the dashed line cut at $B_1=-0.18~\mathrm{V}$ in Fig.~\ref{fig:stability_diagram_for_line_cuts}. Four peaks between $P_2 = -0.5$ and $-0.4~\mathrm{V}$ arise from the in-situ RF-GD sensing of the reservoir-QD$_2$ charge transitions, while two steps within the broad peak separate three charge states detected by the RF-GD-SHB sensor. Figures~\ref{fig:dependence}(a,b) show the temperature dependence of the resonator amplitude and phase response, respectively. The charge transition for the auxiliary QD, near $n_2=0 \rm~{or}~1$, that shifts with DQD occupation, decreases by a factor of seven, as mixing chamber temperature increases from 10 mK to 300 mK, while width of the peak decreases by much less, as expected because the lowest reservoir temperature is $\approx 150~\mathrm{mK}$. As dielectric loss experienced by the circuit capacitance decreases (dissipation increases), the RF-GD circuit moves further away from the impedance-matching condition. We rule out thermal quasiparticle-induced losses in the NbTiN at our modest $Q\approx 10$ ($T_C\approx 11$~K) using control experiments on Si substrates. Potential sources of  dielectric loss are plentiful: imperfections in the $\mathrm{Al_2O_3}$ gate dielectric, the interfaces near the quantum well, the strain relaxation layers, and the PCB, as well as activated conduction in the quantum well \cite{nigroDemonstrationMicrowaveResonators2024}.

We discuss the RF-GD response as a function of excitation power to show that SNR is maximized at approximately $-95$~dBm, the power used in Fig.~\ref{fig:two_modes}. Figures~\ref{fig:dependence}(c,d) show the dependence of the RF-GD response on the applied RF power. At low excitation power (for example, at $n_2=0~\textrm{or}~1$), increasing the drive increases the signal (amplitude and phase) relative to the measurement noise, up to a maximum of $-95$~dBm, after which further increase in drive reduces the signal, similar to previous work on Si\cite{denisovDispersiveReadoutSilicon2023}. Since we do not observe power broadening of the lines in our data, we attribute the reduced signal to degraded matching from higher dielectric loss (smaller $R_\varepsilon$) at higher powers, consistent with Fig.~\ref{fig:setup_and_tank_response}(c,d).

Finally, we investigate the frequency dependence of the RF-GD response, showing that it arises from dissipative effects, not reactive effects, as expected since our parasitic capacitance is large compared to tunneling capacitance. Figures~\ref{fig:dependence}(e,f) show the RF-GD response as a function of probe frequency. For both the $\textrm{QD}_2$ and auxiliary QD loading lines, the largest amplitude contrast occurs near the $f_0=540~\mathrm{MHz}$ LC resonance, as expected for a dissipative mechanism. Similarly, the phase contrast is maximized, and opposite in sign, on the flanks near $530$ and $550~\mathrm{MHz}$. That is, crossing a load line increases (decreases) the angle for $f>f_0$ ($f<f_0$), as expected for a dissipative mechanism.

Using our experimental data, we argue that fast ($\lesssim 1$~$\mu$s) with high-fidelity (99\%), and GD readout is experimentally attainable, with realistic improvements to our device and circuit, and with fewer improvements if the SHB is employed. The long measurement times $t_m$ in our experiment are undesirable for two reasons. First, when $t_m$ increases beyond the interdot transition lifetime (reported as $T_1=17.2\pm3.2~\mu\mathrm{s}$~\cite{rooneyGateModulationHole2025} and $T_1=26~\mu\textrm{s}$~\cite{hendrickxSingleholeSpinQubit2020} in recent experiments), readout fidelity suffers. Second, when $t_m$ increases beyond the decoherence time of neighboring qubits, errors occur in un-measured qubits during mid-circuit measurement, required e.g. in quantum error correction. Our calculation (Appendix~\ref{appendix: comments on high-fidelity single-shot readout}) shows that a reduction in dielectric loss to $R_\varepsilon\approx 100$~k$\Omega$ (by $12\times$) yields a $100\times$ large power signal ($100\times$ power SNR improvement), and that this is achievable, given recent results on superconducting resonators deposited on Ge quantum wells \cite{ruggiero2026high}. A reduction in noise-power reaching $25\times$ has recently been demonstrated over HEMT-based amplification approaches, using parametric amplifiers \cite{fadavi2023three} which can operate in high magnetic fields \cite{xu2023magnetic,khalifa2023nonlinearity}. Achieving a target fidelity of 99~\% requires an improvement of $(21.6/18.5)\times(50~\rm{ms}/1~\mu\rm{s})\approx 60,000$ in SNR, around $24\times$ beyond the compound improvement from reduced loss ($100\times$) and improved amplification ($25\times$). A modest improvement of $2.2\times$ on the lever arm difference $\alpha'$ would make up the remaining $24\times$ increase, due to the $\alpha'^4$ dependence of GD signal. This could likely be achieved with small changes to the device, e.g., plunger gates in the bottom layer, reduced dielectric thickness, or both. Combining this fast charge detection with high fidelity spin-to-charge conversion \cite{kelly2026identification} recently demonstrated should enable the use of compact gate-dispersive readout circuits. Turning to the potential of fast, high-fidelity RF-GD-SHB readout, we note that two improvements are needed. First, the SHB is an unintentional QD in our experiment. Deliberately incorporated SHBs with controlled capacitive coupling to the target QDs could be used, as discussed elsewhere \cite{chittock-woodRadiofrequencyCascadeReadout2026}. Second, a higher SNR will also be required, though only by a factor of $3000\times$ (c.f. $60,000\times$ for RF-GD). Both RF-GD and RF-GD-SHB therefore have the potential to deliver fast, high-fidelity readout.

In conclusion, We have demonstrated RF-GD charge detection in a planar Ge/SiGe DQD at 540 MHz, resolving both QD loading and interdot charge transitions. The in-situ GD detection of the $(1,1)$--$(2,0)$ transition yielded a power-SNR of $18.5\pm3$ for a $50~\textrm{ms}$ integration time, corresponding to a unity-SNR time of $\approx 2.7~\textrm{ms}$. By tuning an auxiliary QD into the SHB regime, the same transition produced a power-SNR of $372\pm12$, a $\sim20$-fold enhancement, reducing the unity-SNR time to $\approx 135~\mu$s. Temperature- and power-dependent show that dielectric loss limits the present sensitivity. Supported by experimental data and a simple RF-GD model, our findings indicate that fast ($\lesssim 1\mu\rm{s}$) and high-fidelity (99\%) GD readout is feasible by reducing dielectric loss, integrating parametric amplification, and improving gate lever arms. These results establish GD sensing, both \textit{in-situ} and SHB, as a compact approach to charge readout in Ge/SiGe.

\begin{acknowledgments}
 The authors acknowledge Eva Dupont-Ferrier, Jonathan Baugh, Jan Kycia, and Sergei Studenikin for helpful feedback. The work at UBC was supported by the Stewart Blusson Quantum Matter Institute (SBQMI) and the Canada First Research Excellence Fund, Quantum Materials and Future Technologies Program. J.S. acknowledges financial support from NSERC (Discovery Grant Program and Quantum Alliance Consortium “Consortium on Quantum Simulation with Spin Qubits”) and from Defence Canada (Innovation for Defence Excellence and Security, IDEaS), the Canada Foundation for Innovation (CFI) John R. Evans Leaders Fund, the CFI Innovation Fund, the NSERC Discovery Grants program, the Pacifican Regional Quantum Initiative, and the Mitacs Accelerate program. M.R.T. acknowledges financial support from the Four Year Doctoral Fellowship (4YF) at UBC. M.R.T., E.S. and J.S. thank all members of the UBC Quantum Science and Technology Laboratory for helpful discussions and comments.
\end{acknowledgments}

\vspace{\baselineskip} 

G.S. and M.V. are founding advisors of Groove Quantum BV and declare equity interests. J.S. and E.S. are co-founders of and hold equity in  Qlithic Technologies. 

\vspace{\baselineskip} 

\appendix
\section{Heterostructure and device fabrication} \label{appendix: Heterostructure and device fabrication}

The device was fabricated on a Ge/SiGe heterostructure grown on a Si wafer. The heterostructure hosts a two-dimensional hole gas (2DHG) in a strained Ge quantum well located approximately 55 nm below the surface \cite{lodariLowPercolationDensity2021}. The buried quantum well provides the conducting channel for the accumulation-mode QDs investigated in this work.

Device fabrication was performed at the University of British Columbia using a combination of electron-beam lithography (EBL), optical lithography, electron-beam metal deposition, atomic layer deposition (ALD), sputtering, wet etching, and liftoff. First, Ti/Au (5/55~nm) alignment markers and ohmic bond pads were defined in a single EBL and liftoff step. A 30-nm-thick Al layer was subsequently deposited and patterned by optical lithography and liftoff to form the ohmic contacts. Immediately prior to ohmic metal deposition, the exposed semiconductor surface was treated with a buffered hydrofluoric acid (BHF) etch to remove the native oxide.

To improve electrical isolation of the gate bond pads, a 220-nm-thick silicon nitride layer was patterned using a combination of optical lithography, silicon nitride deposited by pulsed DC sputtering, and liftoff. This layer prevents the bond pads from being electrically shorted to the quantum well, after wire bonding, which could otherwise result in gate leakage. The first 10-nm-thick Al$_2$O$_3$ dielectric layer was deposited by ALD at 280~$^\circ$C, after the silicon nitride deposition. The ALD includes a 1-hour hold time at 280~$^\circ$C to which assists in forming ohmic contacts between the Al and the Ge quantum well.

The nanoscale (fine) gates were fabricated in two metal layers. The first fine gate layer was defined by EBL,  electron beam deposition of Ti/Pd (3/12~nm), and liftoff. The first fine gate layer consists of screening $(S_i)$ gates and barrier $(B_i \ \textrm{and} \ B_{ij})$ gates discussed in the main text. A second 15-nm-thick Al$_2$O$_3$ layer was deposited by ALD at 280~$^\circ$C, and a second fine gate layer was defined by EBL, electron beam deposition of Ti/Pd (4/16~nm), and liftoff. The second fine gate layer consists of the plunger $(P_i)$ gates and accumulation reservoirs $(Acc_i)$ discussed in the main text.

Following fabrication of the fine gates, BHF etching was used to define vias for electrical connection to the first fine-gate layer and to open the previously fabricated ohmic bond-pad regions for wirebonding. Finally, gate bond pads and macroscopic wiring to the vias were fabricated by optical lithography, electron beam deposition of Ti/Au (25/250~nm) and liftoff. The resulting multilayer gate stack enables independent electrostatic control of the QD potential landscape while maintaining electrical isolation from one layer to another.

\section{Measurement Setup}
\label{appendix: Measurement setup}

Measurements were performed in a dilution refrigerator with a base temperature of approximately $10~\mathrm{mK}$ (BlueFors LD250). Unless otherwise specified, all measurements were performed at the base temperature. The sample was mounted on a custom-designed printed circuit board (PCB), enclosed in a custom-machined oxygen-free copper (OFC) PCB carrier box, and thermally anchored to the mixing-chamber (MXC) plate using a custom OFC cold finger in the bore of the 10~T magnet.

DC gate voltages were generated using a 24-channel digital to analog converter (QDAC-I) and applied to the device through a custom six-stage low-frequency filter with a cutoff of approximately 60~kHz, located at the MXC stage. The DC currents ($I_{SHT}$ and $I_{DQD}$) through the device were measured using low-noise current preamplifiers (DLCPA-200) and digitized by a multi-channel digitizer (National Instruments PCIe-6351).

For reflectometry, the RF excitation was generated using a low phase-noise RF signal generator (Rohde\&Schwarz SMIQ06B) and applied to the gate-tank circuit using a directional coupler (ZEDC-15-2B). The reflected RF signal was amplified using an HEMT-based low-noise amplifier mounted on the 4~K plate (LNF-LNC0.2\_3B) and further amplified at room temperature and subsequently demodulated using a commercially available quadrature demodulator (AD0105B polyphase). The resulting in-phase and quadrature signals were digitized by the same multi-channel digitizer used to acquire the current signals (NI PCIe-6351).

\section{Additional data} \label{appendix: Additional data}

\begin{figure}
    \centering
    \includegraphics{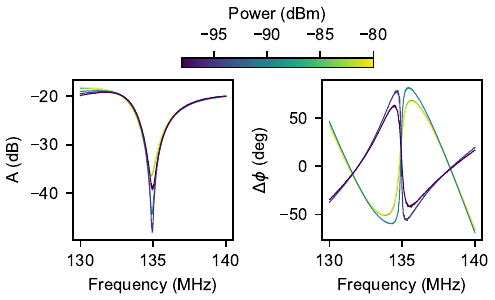}
    \caption{Amplitude and phase response of the RF-SHT sensor when both SHT and DQD are OFF (all gates are at $0 ~\textrm{V}$). A resonance at $\sim135~\textrm{MHz}$ is observed, and it experiences critical matching condition near -95 dBm, indicated by the largest dip in the amplitude $A$ and the sign reversal of the phase change $\Delta \phi$. A linear background is subtracted from all phase plots.}
    \label{fig:RF-SHT characterization}
\end{figure}

Figure~\ref{fig:RF-SHT characterization} shows the amplitude and phase response of the RF-SHT sensor as a function of excitation power. When all the gates are at $0 \textrm{V}$, both the SHT and DQD are off, the resonator exhibits a resonance near $135~\mathrm{MHz}$. Increasing the excitation power from $~-100$ to approximately $-95~\mathrm{dBm}$ improves the impedance matching, resulting in an enhanced resonance dip. At powers above $-95~\mathrm{dBm}$, however, the matching progressively deteriorates with increasing excitation power. We also observe a reversal in the sign of the phase response, $\Delta\phi$, for powers above $-95~\mathrm{dBm}$. Both the amplitude and phase response are consistent with the circuit passing through the optimal matching condition at around -95 dBm. Using $L=700$~nH, we deduce $C\approx1.9$~pF---higher than the GD circuit, partially due to the intentional surface mount capacitor---and an optimally matched condition $R=L/CZ_0 \approx 8$~k$\Omega$. The absence of the screening gate under reservoir accumulation gates is known to yield poor performance of the RF-SET \cite{connorsRapidHighFidelitySpinState2020} and notably, a local screening gate $S_1$ is present on the RF-GD sensor under $P_1$ and absent on RF-SHT sensor under the accumulation gate $\rm{Acc}_3$.

\begin{figure}
    \centering
    \includegraphics{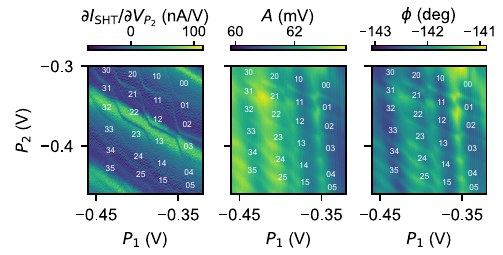}
    \caption{Charge stability diagrams as a function of $P_1-P_2$ gate-space. The SHT is not compensated and therefore we see multiple Coulomb oscillations within the scanned range, and in the Coulomb-blockaded region the SHT is not sensitive to DQD charge transitions. In the gate-tank responses, in addition to the transitions observed in the SHT response, we also see many parallel stripes strongly coupled to $P_1$. The dot-occupation label $nm$ represents $n$ holes in $\textrm{QD}_1$ and $m$ holes in $\textrm{QD}_2$.}
    \label{fig:P1-P2 gate-space}
\end{figure}

\begin{figure}
    \centering
    \includegraphics[width=0.5\textwidth]{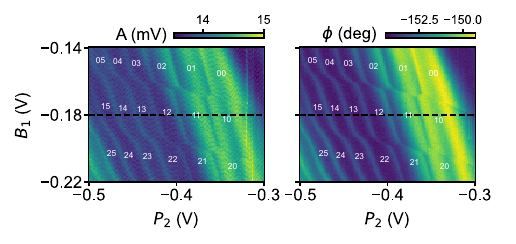}
    \caption{Charge stability diagram indicating the line cut at $B_1=-0.18~\textrm{V}$ used to investigate the dependence of the gate-tank response on temperature, excitation power, and probe frequency. The dot-occupation label $nm$ represents $n$ holes in $\textrm{QD}_1$ and $m$ holes in $\textrm{QD}_2$. Compared with the stability diagram in the first column of Fig.~\ref{fig:moving_stripes}, the measurements were performed at $B_{12}=-0.125~\textrm{V}$ instead of $-0.127~\textrm{V}$, and at $f_{\mathrm{probe}}=550~\textrm{MHz}$ instead of $541.8~\textrm{MHz}$.}
    \label{fig:stability_diagram_for_line_cuts}
\end{figure}

\begin{figure}
    \centering
    \includegraphics[]{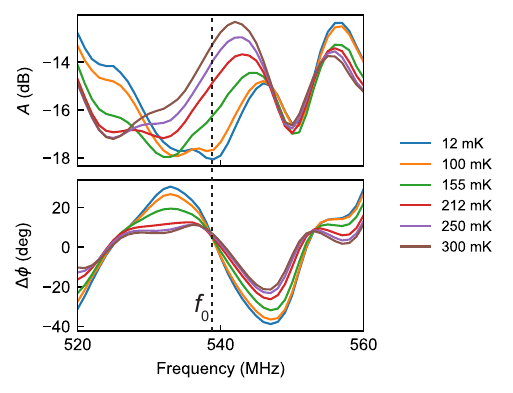}
    \caption{Amplitude and phase response of the gate tank at varying MXC temperature. The DQD is tuned to a few-hole regime and the charge sensor is tuned at the shoulder of a Coulomb peak. The resonance frequency $f_0$ is marked with a hatched vertical line. }
    \label{fig:Temperature_S11}
\end{figure}

\section{Prospects for high fidelity single shot readout}\label{appendix: comments on high-fidelity single-shot readout}

\begin{figure*}
    \centering
    \includegraphics{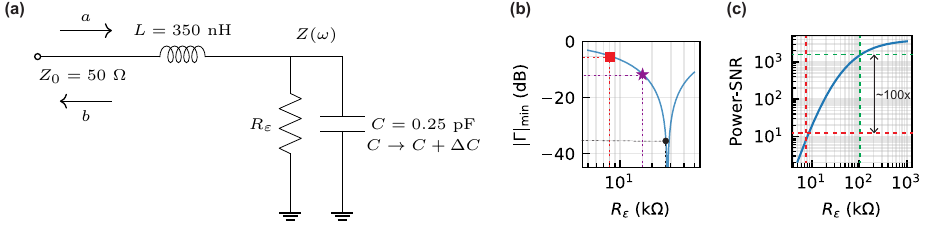}
    \caption{Lumped-element circuit model of the RF GD resonator. (a) The resonator consists of a series inductance $L$ followed by a parallel resistance $R_\varepsilon$---representing dielectric loss---and capacitance $C$. A state-dependent capacitance change $\Delta C$ modifies the complex impedance and therefore the measured reflection coefficient $\Gamma=(Z-Z_0)/(Z+Z_0)$. (b) Minimum reflection coefficient magnitude ($|\Gamma|_\textrm{min}$) as a function of $R_\varepsilon$. The estimated loss resistance is indicated for each operating state of the DQD and SHT (circle: DQD off, SHT off; star: DQD on, SHT off; square: DQD on, SHT on). (c) Power SNR is calculated from $\textrm{SNR}=|\Delta \Gamma|^2 \displaystyle \frac{P_{in}}{P_{N}}$, where noise power ${P_{N}}=k_BT_{sys}B$. For the calculation, $\Delta C=20$~aF, system noise temperature $T_{sys}=4~\textrm{K}$, measurement bandwidth $B$, and input power $P_{in}=-95~\textrm{dBm}$ are assumed.  The SNR is evaluated at the frequency where the phase change is maximum. The SNR improvement discussed in the main text is a ratio of two SNRs, so it is independent of quantum capacitance.}
    \label{fig:CircuitModel_SNR}
\end{figure*}

The GD readout demonstrated in the main text provides a baseline from which to assess the potential of fast, single-shot, high-fidelity readout in planar Ge/SiGe QDs. We choose a target of $99\%$ in a measurement time less than $1~\mu\mathrm{s}$. We make the following assumptions: (1) the noise is white and Gaussian, (2) the readout is fast compared to the relaxation time at the readout point, and (3) an ideal spin-to-charge mapping with no infidelity. Recently, high-fidelity (97~\%) spin-to-charge conversion and readout has been demonstrated\cite{kelly2026identification} with 50 microsecond integration, limited by SNR of the RF-SHT, and 2 microsecond spin-to-charge conversion, and fast $100$~ns spin-to-charge conversion has been obtained in experiments by modulating the tunnel barrier of Si/SiGe DQD during readout\cite{takedaRapidSingleshotParity2024}. 

Our approach is as follows. First, we extrapolate the required enhancement of SNR to perform readout in the desired measurement time, using the measurement SNR and measurement time from experiments. Second, we will explore physical mechanisms to achieve the desired SNR enhancement. 

For a 99\% fidelity measurement of two Gaussian-distributed measurement outcomes with equal variance, the required SNR is $21.6$. The experiment in the main text achieves an SNR of $18.5$ for a measurement time of $50~\mathrm{ms}$. Since SNR scales linearly with integration time, assuming the noise is white, the target SNR in $1~\mu\mathrm{s}$ requires a substantial SNR improvement of $21.6\times 50\textrm{ ms}/(18.5\times 1\textrm{ }\mu\textrm{s})\approx 6\times10^4$.

Below we argue that a power SNR improvement of up to $100\times25\times81\approx 2\times10^5$, exceeding the required improvement of $6\times10^4$, can be obtained from (i) a $100\times$ SNR improvement from reduction in dielectric loss, (ii) a $25\times$ SNR improvement from switching to quantum-limited amplification, and (iii) a $81\times$ improvement obtained from a modest $3\times$ increase in lever arm.  \\

\noindent {\bf (i) Dielectric Loss}
A reduction in dielectric loss to $100$~k$\Omega$ can improve power SNR by around $100$, and is achievable with currently available Ge quantum well materials. We estimate $R_\varepsilon\approx 25$~k$\Omega$, $R_\varepsilon\approx 15$~k$\Omega$, and $R_\varepsilon\approx 8$~k$\Omega$ for no accumulation, accumulation on the DQD, and accumulation on the SHT and QDs [see Fig.~\ref{fig:CircuitModel_SNR}(b)], as argued in the main text. Figure ~\ref{fig:CircuitModel_SNR}(c) shows the dependence of SNR in our experiment on dielectric loss. A target loss of $R_\varepsilon\approx 100$~k$\Omega$ increases power-SNR by $100\times$ c.f. $R_\varepsilon\approx 8$~k$\Omega$ [see Fig.~\ref{fig:CircuitModel_SNR}(c)]. Notably, it should be possible to realise a dielectric loss resistance exceeding $R_\varepsilon \approx 100$~k$\Omega$ with Ge substrates available today. Recently, 5.86~GHz superconducting coplanar waveguide resonators with internal quality factors well above $10,000$ have been realized on Ge quantum well substrates\cite{ruggiero2026high}. The dielectric loss resistance associated with $Q_i\approx 10000$ is $R_\varepsilon=\pi Q_i Z_0 / 2\approx 1$~M$\Omega$. \\

\noindent {\bf (ii) Quantum-limited amplification}
An enhancement of power SNR by $25$ is possible by reducing the system noise temperature using a quantum-limited amplifier operated at millikelvin temperature. An enhancement of $10$ dB is common, and enhancements up to $12$ dB to $14$ dB have recently been obtained  \cite{fadavi2023three}. \\

\noindent {\bf (iii) Lever arm}
Finally, increasing the gate-to-dot lever arm provides a direct route to enhancing the dispersive response. A threefold increase in lever arm is estimated to provide an approximately $3^4=81\times$ ($\textrm{SNR}\propto |\Delta \Gamma|^2$ and $|\Delta \Gamma|\propto \alpha'^2$) improvement in SNR. Such an increase in $\alpha'$ is experimentally plausible in the present device geometry. The RF tank is coupled through the top plunger layer, while the bottom barrier layer has a lever arm approximately twice that of the plunger gates. Reducing the dielectric thickness and connecting the RF tank to an electrode in the bottom layer would therefore bring the sensing electrode closer to the QD and increase its capacitive coupling. Further enhancement could be achieved by modestly increasing the electrode area. These geometric modifications could reasonably provide a threefold enhancement in lever arm without requiring a fundamentally different device architecture.\\

\noindent{\bf Input power}
Although the SNR increases linearly with input power, it would be difficult to increase the probe RF power further because, even at $-95~\textrm{dBm}$, the estimated detuning modulation amplitude is approximately $11~\mu\textrm{eV}$ (including the voltage enhancement by the resonator) with the above proposed improvements in dielectric loss and gate lever arm. This is comparable to the interdot tunnel coupling, $t_c \approx 20.7~\mu\textrm{eV}$ ($t_c/h=5~\textrm{GHz}$). Increasing drive power beyond the tunnel coupling does not increase the signal appreciably.

\section{Extraction of the interdot tunnel coupling in direct regime} \label{appendix: tunnel coupling in direct regime}

The interdot tunnel coupling is extracted from the GD phase response near the interdot charge transition. We model the DQD as a two-level system formed by the charge states on either side of the transition. In the vicinity of the transition, the Hamiltonian can be written as
\begin{equation}
H =
\frac{1}{2}
\begin{pmatrix}
\epsilon & \Delta_c \\
\Delta_c & -\epsilon
\end{pmatrix},
\end{equation}
where $\epsilon$ is the detuning energy between the two charge configurations and $\Delta_c$ is related to tunnel coupling $t_c$ by
\begin{equation}
\Delta_c = 2 t_c.
\end{equation}
The resulting energy splitting between the hybridized states is
\begin{equation}
\Delta E =
\sqrt{\epsilon^2+\Delta_c^2}.
\end{equation}

The GD response is related to the curvature of the ground-state energy with respect to the electrostatic potential applied to the gate. For the two-level system considered here, the corresponding quantum capacitance~\cite{mizutaQuantumTunnelingCapacitance2017} at finite temperature is
\begin{equation}
C_Q(\epsilon)
=
\frac{(e\alpha')^2}{2}
\frac{(\Delta_c)^2}{(\Delta E)^3}
\tanh\left(
\frac{\Delta E}{2k_{\mathrm B}T}
\right),
\label{eq:quantum_capacitance_fit}
\end{equation}
where $e$ is the elementary charge, $k_{\mathrm B}$ is the Boltzmann constant, $T$ is the hole temperature, and $\alpha'=\alpha_1 - \alpha_2$ is the difference in lever arms of gate $P_1$ to $\textrm{QD}_1$ and $\textrm{QD}_2$.

For the tunnel-coupling extraction, the temperature is fixed at $T=100~\mathrm{mK}$. The measured phase response is assumed to be proportional to the quantum capacitance; and a linear background is included to account for the slowly varying phase response unrelated to the interdot transition. The fitting function is therefore
\begin{equation}
\phi(\epsilon)
=
A C_Q(\epsilon-\epsilon_0)
+
m(\epsilon-\epsilon_0)
+
\phi_0,
\label{eq:phase_fit}
\end{equation}
where $A$ is an overall conversion factor between quantum capacitance and measured phase, $\epsilon_0$ accounts for an offset in the detuning coordinate, $m$ describes a linear background, and $\phi_0$ is a constant phase offset. The parameters $t_c$, $\epsilon_0$, $A$, $m$, and $\phi_0$ are obtained simultaneously from a nonlinear least-squares fit to the measured phase response. The uncertainty in $t_c$ is obtained from the covariance matrix returned by the nonlinear least-squares fit. 

\section{Extraction of the interdot tunnel coupling in auxiliary-dot-assisted regime} \label{appendix: tunnel coupling in SHB regime}
In the auxiliary-dot-assisted regime, the phase response associated with the same interdot transition exhibits a step-like feature resembling the response of a dispersive single-electron-box charge sensor. In this regime, the measured signal is modeled in terms of the occupation of the hybridized DQD states rather than directly in terms of the quantum capacitance. The same two-level Hamiltonian introduced above is used. At finite temperature, the excess charge~\cite{dicarloDifferentialChargeSensing2004} occupation of the $\textrm{QD}_2$ site is then written as
\begin{equation}
n_2(\epsilon)
=
\frac{1}{2}
\left[
1+
\frac{\epsilon}{\Delta E}
\tanh\left(
\frac{\Delta E}{2k_{\mathrm B}T}
\right)
\right].
\label{eq:occupation_aux}
\end{equation}
Here, $n_2$ denotes the probability of finding an excess charge in $\textrm{QD}_2$ within the two-level subspace. The finite tunnel coupling smooths the transition between the two charge configurations, while the finite temperature provides an additional broadening of the occupation step.

The measured phase response is assumed to be proportional to the occupation of $\textrm{QD}_2$. We therefore use the model
\begin{equation}
\phi(\epsilon)
=
A n_2(\epsilon-\epsilon_0)
+
m(\epsilon-\epsilon_0)
+
\phi_0,
\label{eq:aux_phase_fit}
\end{equation}

where $A$ is the conversion factor between the charge occupation and the measured phase, $\epsilon_0$ accounts for an offset of the detuning coordinate, $m$ describes a linear background, and $\phi_0$ is a constant phase offset. The fit parameters are $t_c$, $\epsilon_0$, $A$, $m$, and $\phi_0$.

The temperature is again fixed at $T=100~\mathrm{mK}$. The resulting occupation-based model produces a smooth step in the measured phase as the detuning is swept through the interdot transition. In contrast to the quantum-capacitance response, which is proportional to the curvature of the hybridized energy spectrum and therefore produces a peak-like feature centered near zero detuning, the occupation response changes monotonically across the transition. The width of this transition is determined by both the tunnel coupling and the thermal energy. Consequently, fitting the step-like phase response provides an estimate of the tunnel coupling $t_c$. The tunnel coupling and its uncertainty are obtained from a nonlinear least-squares fit to Eq.~\eqref{eq:aux_phase_fit}.

\section{Constant-interaction model for Direct, Auxiliarry-dot-assisted: SHB, and Auxiliarry-dot-assisted: Cascade regimes} \label{appendix: Constant-interaction model}
\begin{figure*}
    \centering
    \includegraphics[width=0.75\linewidth]{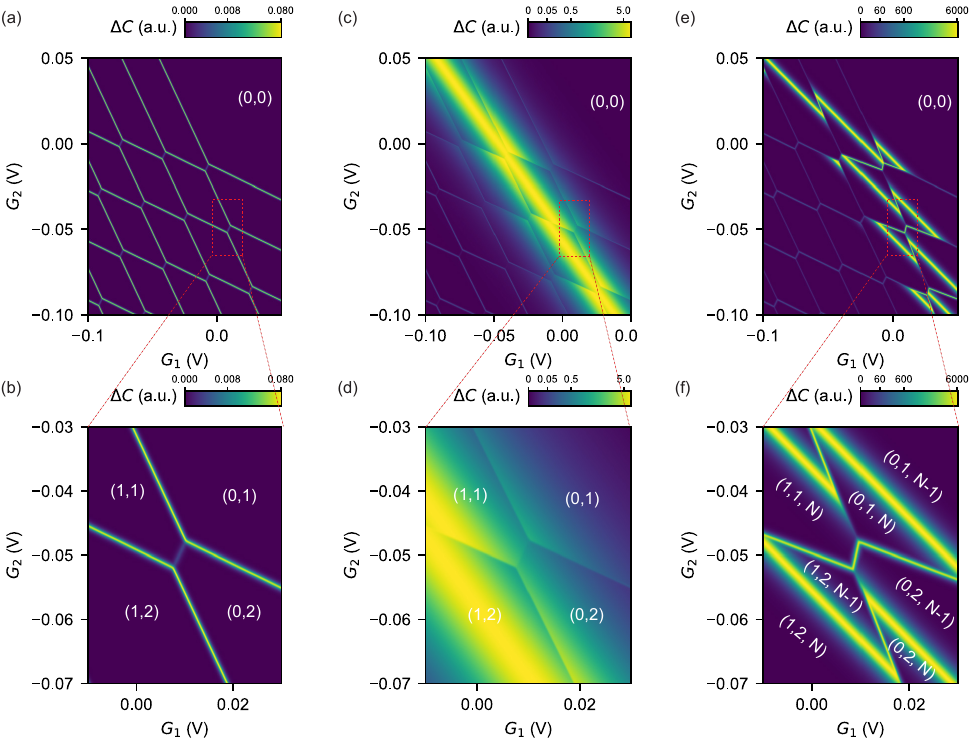}
    \caption{Simulated charge stability diagrams obtained from the constant-interaction model for the three charge-readout regimes. (a) Charge stability diagram in direct GD regime over a large gate-voltage range, showing the load lines of dots~1 and~2 and the interdot charge-transition lines. (b) Enlarged view of the region near the $(1,1)$--$(0,2)$. (c) Charge stability diagram in a regime where auxiliary-dot works as an SHB, showing the additional load lines associated with the charge transitions of the auxiliary dot. (d) Enlarged view near the $(1,1)$--$(0,2)$ transition in auxiliary-dot-assisted SHB regime. (e) Charge stability diagram in which the auxiliary-dot facilitates the cascade process, where the stronger mutual Coulomb interactions between the auxiliary-dot and the DQD produce correlated charge-transition features. (f) Enlarged view near the $(1,1)$--$(0,2)$ transition in auxiliary-dot-assisted cascade regime, highlighting the cascade response associated with the auxiliary-dot charge transitions from $N-1$ to $N$.}
    \label{fig:Simulated_data}
\end{figure*}
\begin{table*}[t]
\centering
\caption{Parameters used in the constant-interaction simulations for the three charge-readout regimes. The temperature is $T=100$~mK for all simulations. The interaction-energy matrix is defined by $\mathbf{U}=e^2\mathbf{C_{\mathrm{cc}}}^{-1}$, and the entries are given in meV. The gate-to-dot capacitance matrix $\mathbf{C_{\mathrm{cv}}}$ is given in aF.}
\label{tab:simulation_parameters}
\begin{ruledtabular}
\begin{tabular}{l c c c}
Regime &
$n_3$ &
$\mathbf{U}$ (meV) &
$\mathbf{C_{\mathrm{cv}}}$ (aF)
\\
\hline

Direct GD; Figs.~\ref{fig:Simulated_data}(a, b) &
0 &
$\begin{pmatrix}
2.50 & 0.45 & 0.008\\
0.45 & 2.60 & 0.010\\
0.008 & 0.010 & 0.45
\end{pmatrix}$ &
$\begin{pmatrix}
5.34 & 2.54 & 0.08\\
1.07 & 5.00 & 0.06\\
0.60 & 0.50 & 5.20
\end{pmatrix}$
\\[18pt]

Auxiliarry-dot-assisted: SHB; Figs.~\ref{fig:Simulated_data}(c, d) &
{0, 1, 2, ...} &
$\begin{pmatrix}
2.50 & 0.45 & 0.008\\
0.45 & 2.60 & 0.010\\
0.008 & 0.010 & 0.45
\end{pmatrix}$ &
$\begin{pmatrix}
5.34 & 2.54 & 0.08\\
1.07 & 5.00 & 0.06\\
0.60 & 0.50 & 5.20
\end{pmatrix}$
\\[18pt]

Auxiliary-dot-assisted: Cascade; Figs.~\ref{fig:Simulated_data}(e, f) &
{0, 1, 2, ...} &
$\begin{pmatrix}
2.50 & 0.45 & 0.12\\
0.45 & 2.60 & 0.30\\
0.12 & 0.30 & 2.45
\end{pmatrix}$ &
$\begin{pmatrix}
5.34 & 2.54 & 0.08\\
1.07 & 5.00 & 0.06\\
0.40 & 0.30 & 30.2
\end{pmatrix}$
\\

\end{tabular}
\end{ruledtabular}
\end{table*}
We model the charge stability of the three-dot system using a constant-interaction approximation. For a charge configuration $\mathbf{n}=(n_1,n_2,n_3)^T$, where $n_i\in\mathbb{Z}_{\geq 0}$ denotes the integer number of holes occupying dot $i$, the electrostatic energy~\cite{vanderwielElectronTransportDouble2002} is
\begin{equation}
E(\mathbf{n},\mathbf{V}_g)
=
\frac{e^2}{2}
(\mathbf{n}-\mathbf{n}_g)^T
\mathbf{C_{\mathrm{cc}}}^{-1}
(\mathbf{n}-\mathbf{n}_g),
\end{equation}
where $\mathbf{C_{\mathrm{cc}}}$ is the dot-to-dot capacitance matrix and the gate-induced charge is
\begin{equation}
\mathbf{n}_g
=
-\frac{\mathbf{C_{\mathrm{cv}}}\mathbf{V}_g}{e}.
\end{equation}
The negative sign accounts for our convention that increasingly negative gate voltages accumulate holes. The corresponding interaction-energy matrix is
\begin{equation}
\mathbf{U}=e^2 \mathbf{C_{\mathrm{cc}}}^{-1},
\end{equation}
where the diagonal elements describe the charging energies of the individual dots and the off-diagonal elements describe their mutual electrostatic interactions.

At finite temperature, the probability of a given charge configuration is obtained from the Boltzmann distribution,
\begin{equation}
P(\mathbf{n})
=
\frac{\exp[-E(\mathbf{n})/k_{\mathrm{B}}T]}
{\sum_{\mathbf{n}'}
\exp[-E(\mathbf{n}')/k_{\mathrm{B}}T]},
\end{equation}
from which the average occupation of each dot is calculated as
\begin{equation}
\langle n_i\rangle
=
\sum_{\mathbf{n}} n_i P(\mathbf{n}).
\end{equation}

The differential capacitance~\cite{mizutaQuantumTunnelingCapacitance2017}, seen by the gate $G_3$, is calculated using
\begin{equation}
\Delta C
=
\frac{\partial \mathbf{Q}}{\partial V_{G_3}}
=e\sum_i
\alpha_{i3}
\frac{\partial\langle n_i\rangle}{\partial V_{G_3}},
\end{equation}
where the lever arm matrix $\boldsymbol{\alpha}$ is defined as
\begin{equation}
\boldsymbol{\alpha}
=
\frac{1}{e}(\mathbf{C_{\mathrm{cc}}}^{-1}\mathbf{C_{\mathrm{cv}}}).
\end{equation}

Three regimes are considered. First, when the auxiliary dot remains fixed at $n_3=0$, shown in Figs.~\ref{fig:Simulated_data}(a,b), only the load lines of dots 1 and 2 and the interdot charge-transition lines are present. Second, when the auxiliary dot is also allowed to change occupation, additional broad load lines associated with $n_3\rightarrow n_3+1$ appear, as shown in Figs.~\ref{fig:Simulated_data}(c,d). This regime corresponds to operation of the auxiliary dot as a single-hole-box charge sensor. Finally, when the mutual interactions $U_{13}$ and $U_{23}$ are sufficiently strong, loading either DQD dot or an interdot transition produces a substantial shift of the electrochemical potential of the auxiliary dot,
\begin{equation}
\Delta\mu_3
\simeq
U_{31}\Delta n_1
+
U_{32}\Delta n_2.
\end{equation}
When this shift exceeds $3.5k_{\mathrm{B}}T$, a DQD transition can trigger a change in the auxiliary-dot occupation, resulting in correlated charge transitions and the cascade effect~\cite{vandiepenElectronCascadeDistant2021, chittock-woodRadiofrequencyCascadeReadout2026}, as shown in Figs.~\ref{fig:Simulated_data}(e,f).

%

\end{document}